\documentclass[journal,twocolumn]{IEEEtran}

\usepackage{multicol}
\usepackage{etoolbox}
\makeatletter
\patchcmd{\@makecaption}
  {\scshape}
  {}
  {}
  {}
\makeatletter
\patchcmd{\@makecaption}
  {\\}
  {.\ }
  {}
  {}
\makeatother

\usepackage{amsfonts}
\usepackage{mathrsfs}
\usepackage{mathtools}
\usepackage{amsfonts}
\usepackage{amssymb}
\usepackage{graphicx}
\usepackage{cite,graphicx,amsmath,amssymb,color}
\usepackage{epsfig}
\usepackage{psfrag}
\usepackage{amsmath}
\usepackage{array}
\usepackage{cases}
\usepackage{eufrak}

\usepackage{algorithmic}
\usepackage{algorithm}
\usepackage{subfigure}
\usepackage{bm}
\usepackage{multirow}
\usepackage{threeparttable}
\usepackage{array}
\usepackage{makecell}

\DeclareMathOperator*{\argmax}{argmax}

\newtheorem{Lem}{Lemma}

\newtheorem{Prob}{Problem}

\makeatletter
\renewcommand\normalsize{%
   \@setfontsize\normalsize\@xpt\@xiipt
   \abovedisplayskip 0.025\p@ \@plus0.05\p@ \@minus0.125\p@
   \abovedisplayshortskip \z@ \@plus0.075\p@
   \belowdisplayshortskip 0.15\p@ \@plus0.075\p@ \@minus0.025\p@
   \belowdisplayskip \abovedisplayskip
   \let\@listi\@listI}
\makeatother

\IEEEoverridecommandlockouts

\begin{document}
\vspace{-1cm}
\title{Optimal Multicast of Tiled 360 VR Video in OFDMA Systems}
\vspace{-1cm}
\author{Chengjun Guo\thanks{Manuscript received September 2, 2018; accepted September 21,
2018. The work of Y. Cui was supported by NSFC grant 61401272 and grant 61521062.
The work of Z. Liu was supported  by JSPS KAKENHI Grant Number JP16H02817 and JP18K18036.
The associate editor coordinating the review of this paper and approving it for publication was K. Psannis. \textit{(Corresponding author: Ying Cui.)}
\newline{\indent C. Guo and Y. Cui are with the Institute of Wireless Communication Technologies, Shanghai Institute for Advanced
Communication and Data Science, Department of Electronic
Engineering, Shanghai Jiao Tong University, Shanghai 200240, China (e-mails: \{guochengjun382, cuiying\}@sjtu.edu.cn).}
\newline{\indent Z. Liu is with Department of Mathematical and Systems Engineering, Shizuoka University, Hamamatsu 432-8561, Japan (e-mail: liu@ieee.org).}
},  Ying Cui,~\IEEEmembership{Member,~IEEE}, and Zhi Liu,~\IEEEmembership{Member,~IEEE}}
\vspace{-1cm}
\maketitle
\vspace{-1cm}
%%%%%%%%%%%%%%%%%%%%%%%%%%%%%%%%%%%%%%%
\begin{abstract}
In this letter, we study optimal  multicast of tiled 360 virtual reality (VR) video
from one \textcolor{black}{server (base station or access point)} to multiple users \textcolor{black}{in an orthogonal frequency division multiple access (OFDMA) system.}
For given video quality, we optimize the \textcolor{black}{subcarrier, transmission power and transmission rate} allocation  to minimize the \textcolor{black}{total} transmission power. For given transmission power budget, we optimize the \textcolor{black}{subcarrier, transmission power and transmission rate  allocation}  to maximize the received video quality. These two optimization problems are  non-convex problems.
\textcolor{black}{We obtain a globally optimal closed-form solution and a near optimal solution of the two problems, separately,}  \textcolor{black}{both revealing} important design insights for multicast of tiled 360 VR video \textcolor{black}{in  OFDMA systems}.
\end{abstract}

\section{Introduction}

A  \textit{virtual reality} (VR) video is generated by capturing a scene of interest from all directions at the same time. A user wearing a VR headset  can freely watch the scene of interest in any viewing direction at any time, hence enjoying immersive viewing experience. VR has vast applications in entertainment, education, medicine, etc.
Transmitting an entire 360 VR video which is of a much larger size than a traditional video brings a heavy burden to wireless networks.
To improve transmission efficiency and avoid view switch delay, a 360 VR video is divided into smaller rectangular segments of the same size, referred to as tiles, and the set of tiles covering a user's current \textit{field-of-view} (FoV) and the FoVs that may be watched shortly should be transmitted simultaneously \textcolor{black}{\cite{7532381}}.

\textcolor{black}{In \cite{unicast-liu,unicast-one,ICC2018,multicast-nojoint}, the authors consider 360 VR video transmission in single-user \cite{unicast-liu,unicast-one} and multi-user \cite{ICC2018,multicast-nojoint} wireless networks, and optimize video encoding parameters \cite{unicast-liu,unicast-one} as well as resource allocation \cite{ICC2018,multicast-nojoint} to maximize  \textcolor{black}{the received 360 VR video quality}. The optimization problems in \cite{unicast-liu,unicast-one,ICC2018,multicast-nojoint} are discrete, and the obtained solutions do not offer many design insights.
\textcolor{black}{In \cite{ICC2018},
multicast opportunities are ignored,} and hence the resulting solution may not be efficient for multi-user wireless networks. In \cite{multicast-nojoint}, multicast opportunities are considered, \textcolor{black}{but the tiles are treated} separately in the optimization.
\textcolor{black}{This leads to prohibitively high computational complexity,
as the number of tiles to be transmitted is usually quite large}.
Therefore, \textcolor{black}{it is still not known how the required FoVs and channel conditions of all users affect optimal resource allocation and how to obtain low-complexity resource allocation for 360 VR video transmission  in multi-user wireless networks.}}

In this letter, we would like to address the \textcolor{black}{aforementioned} issues.
We \textcolor{black}{aim to design}  optimal  multicast of tiled 360 VR video from one
 \textcolor{black}{server (base station or access point)} to multiple users \textcolor{black}{in an orthogonal frequency division multiple access (OFDMA) system.}
\textcolor{black}{We formulate
\textcolor{black}{optimal multicast of tiled 360 VR video as multi-group multicast optimization problems.}
Specifically, for}
given video quality, we optimize the \textcolor{black}{subcarrier, power and rate} allocation  to minimize the total transmission power.
\textcolor{black}{We obtain a globally optimal closed-form solution of this problem (under a mild condition), which reveals that
 the minimum transmission power  increases exponentially with the total number of tiles that need to be transmitted.}
\textcolor{black}{For given transmission power budget, we optimize the \textcolor{black}{subcarrier, power and rate} allocation to maximize the received video quality.
We obtain a near optimal solution of this problem, which reveals that the maximum \textcolor{black}{video quality} is inversely proportional to the maximum number of tiles that need to be transmitted for all  viewing directions.}
To the best of our knowledge, these important design insights \textcolor{black}{have never been  analytically verified in existing literature.}
Finally, numerical results demonstrate the advantage of the \textcolor{black}{proposed  solutions}.

\vspace{-0.35cm}
\section{System Model}
\vspace{-0.1cm}

\begin{figure}[t]
\vspace*{-1cm}
\begin{center}
 \includegraphics[width=2.7in]{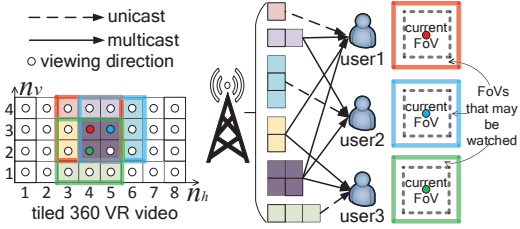}
  \end{center}
  \vspace*{-0.35cm}
     \caption{\small{\textcolor{black}{Illustration} of tiled 360 VR video multicast with $K=3$, $M_h\times\ M_v=8\times4$, $V_h\times V_v=8\times4$.}}  %Note that the user groups are formed according to their quality requirements in Section~\ref{section:power minimization} and according to their distances to the BS in Section~\ref{section:layer selection}, respectively.}}
\label{fig:system-model}
\vspace*{-0.75cm}
\end{figure}

As illustrated in Fig.~\ref{fig:system-model}, we consider  \textcolor{black}{multicast} of a 360 VR video from a single-antenna \textcolor{black}{server (base station or access point)} to $K~(\geq1)$ single-antenna users, \textcolor{black}{denoted by $\mathcal{K}\triangleq\{1,\ldots,K\}$,}
\textcolor{black}{in an OFDMA system}.\footnote{\textcolor{black}{Note that the setup is similar to that considered in our previous
work \cite{VR-TDMA}, except that in this paper, we consider an OFDMA system. We
present the details of the setup here for completeness.}}
\textcolor{black}{Consider} $M_h\times M_v$  viewing directions, where $M_h$ and $M_v$ represent the numbers of horizontal and vertical viewing directions.
The $(m_h,m_v)$-th viewing direction refers to the viewing direction in the $m_h$-th row and \textcolor{black}{$m_v$-th} column. When a  VR user is interested in one viewing direction, he can \textcolor{black}{view} a rectangular FoV of size $F_h\times F_v$ \textcolor{black}{(in rad$\times$rad)}  with the viewing direction as its center.
\textcolor{black}{The viewing direction of each user can be captured by sensors in
his VR headset.}

To improve transmission efficiency, we consider tiling. In particular, the 360 VR video is divided into $V_h\times V_v$ rectangular segments of the same size, referred to as tiles, where $V_h$ and $V_v$ represent the numbers of segments in each row and each column, respectively.
\textcolor{black}{We consider that all tiles have the same encoding rate, denoted by
$D$ (in bit/s).}
\textcolor{black}{Note that the encoding rate reflects the video quality.}
To avoid view switch delay, for each user, the set of tiles that cover its current FoV and the FoVs that may be watched shortly will be delivered.\footnote{\textcolor{black}{
\textcolor{black}{Note that} the proposed framework does not rely on any particular
method for determining the set of tiles to be transmitted to each user \textcolor{black}{\cite{7532381}}.}}

 Let
$\mathbf{X}_k\in\mathcal X$ denote \textcolor{black}{the  viewing direction} of user $k$, where \textcolor{black}{$\mathcal{X}\triangleq\{(m_h,m_v)|m_h=1,\ldots,M_h,m_v=1,\ldots,M_v\}$} represents the set of all possible viewing directions of each user.
Let $\mathbf X\triangleq (\mathbf{X}_k)_{k\in\mathcal K}\in\bm{\mathcal{X}}\triangleq\mathcal X^K$ denote \textcolor{black}{the system viewing direction state,}
\textcolor{black}{and let
$\Phi(\mathbf{X})$ denote the corresponding set of tiles that need to be transmitted to all users.}
In order to \textcolor{black}{utilize multicast opportunities for improving
transmission efficiency,}
 we divide $\Phi(\mathbf{X})$ into $I(\mathbf{X})$  disjoint non-empty sets $\mathcal{S}_i(\mathbf{X})$, \textcolor{black}{$i\in\mathcal{I}(\mathbf{X})\triangleq\{1,\ldots,I(\mathbf{X})\}$.}
For all $i,j\in I(\mathbf{X}),$ $i\neq j$, $\mathcal{S}_i(\mathbf{X})$ and $\mathcal{S}_j(\mathbf{X})$ are for different groups of users.
Let $S_i(\mathbf{X})$ denote the number of tiles in set $\mathcal{S}_i(\mathbf{X})$.
\textcolor{black}{We jointly consider the tiles in each set, instead of treating
them separately (as in \cite{multicast-nojoint}).}\footnote{\textcolor{black}{This
 will lead  to a great
computational complexity reduction for optimal multicast of tiled 360 VR video, compared with \cite{multicast-nojoint}.}}
\textcolor{black}{Let $\mathcal K_{i}(\mathbf X)$  denote the set of users that need to receive the tiles in $\mathcal{S}_i(\mathbf{X})$.}
\textcolor{black}{Then, multicast of tiled 360 VR video can be viewed as multi-group multicast.\footnote{\textcolor{black}{Note that the proposed optimization framework is general
and can be applied to optimally multicast multiple messages
to different groups of users, i.e., multi-group multicast, in OFDMA systems.}}}

\textcolor{black}{Consider $N$ subcarriers,
denoted by $\mathcal{N}\triangleq\{1,\ldots,N\}$.}
Each subcarrier has a bandwidth $B$ (in Hz). \textcolor{black}{Consider one frame.}
Assume block fading, i.e., \textcolor{black}{the channel condition on each subcarrier  does not change within one frame \cite{ICC2018}}.
Let $H_{n,k}\in\mathcal{H}$ denote the power of the channel \textcolor{black}{(i.e., channel state)} on subcarrier $n$ between the \textcolor{black}{server} and
user $k$, where $\mathcal{H}$ denotes the finite channel state space.
Let $\mathbf H\triangleq (H_{n,k})_{n\in\mathcal{N},k\in\mathcal K}\in\bm{\mathcal{H}}\triangleq\mathcal H^{NK}$ denote \textcolor{black}{the  system channel state.}
\textcolor{black}{The  system state consists of   $\mathbf X$ and  $\mathbf H$}, denoted by $(\mathbf X, \mathbf H)\in  \bm{\mathcal{X}}\times\bm{\mathcal{H}}$.
We \textcolor{black}{assume that the server} is aware of the system state $(\mathbf X,\mathbf H)$, \textcolor{black}{e.g., by explicit \textcolor{black}{feedbacks} from all users.}

Let $\mu_{n,i}(\mathbf{X},\mathbf{H})\in\{0,1\}$  denote the subcarrier assignment indicator
\textcolor{black}{for subcarrier $n$ and the tiles in $\mathcal{S}_i(\mathbf{X})$,}
where $\mu_{n,i}(\mathbf{X},\mathbf{H})=1$ indicates that subcarrier $n$ is assigned to transmit the symbols for \textcolor{black}{the} tiles in $\mathcal{S}_i(\mathbf{X})$, and $\mu_{n,i}(\mathbf{X},\mathbf{H})=0$ otherwise.
\textcolor{black}{For ease of implementation, assume} each subcarrier is assigned to transmit symbols \textcolor{black}{for only} one set of tiles.
 Thus, we have the following subcarrier allocation constraints:
\begin{align}
    &\mu_{n,i}(\mathbf{X},\mathbf{H})\in\{0,1\},~i\in\mathcal{I}(\mathbf{X}),n\in\mathcal{N},\label{const:ofdma-integer1}\\
    &\sum\nolimits_{i\in\mathcal{I}(\mathbf{X})}\mu_{n,i}(\mathbf{X},\mathbf{H})=1,~n\in\mathcal{N}.\label{const:ofdma-integer2}
\end{align}

Let $p_{n,i}(\mathbf{X},\mathbf{H})$ \textcolor{black}{and $c_{n,i}(\mathbf{X},\mathbf{H})$   denote the transmission power and transmission rate for}  the symbols for the tiles in $\mathcal{S}_i(\mathbf{X})$  on  subcarrier $n$, respectively, where
\begin{align}
&p_{n,i}(\mathbf{X},\mathbf{H})\geq0,~i\in\mathcal{I}(\mathbf{X}),n\in\mathcal{N},\label{const:ofdma-power1}
\end{align}
\begin{align}
&c_{n,i}(\mathbf{X},\mathbf{H})\geq0,~i\in\mathcal{I}(\mathbf{X}),n\in\mathcal{N}.\label{const:ofdma-power0}
\end{align}
Thus, the \textcolor{black}{total transmission power} is
%\begin{align}\label{eqn:ofdma-total power}
$P(\bm{\mu}(\mathbf{X},\mathbf{H}),\bm{p}(\mathbf{X},\mathbf{H}))=
\sum_{n\in\mathcal{N}}\sum_{i\in\mathcal{I}(\mathbf{X})}\mu_{n,i}(\mathbf{X},\mathbf{H})p_{n,i}(\mathbf{X},\mathbf{H})$.
%\end{align}
To obtain design insights, we consider capacity achieving code.
\textcolor{black}{
Consequently, to guarantee that all users in $\mathcal{K}_{i}(\mathbf{X})$ can successfully receive the tiles in $\mathcal{S}_i(\mathbf{X})$, we have the following transmission rate constraints:
\begin{align}
&\sum\nolimits_{n\in\mathcal{N}}\frac{\mu_{n,i}(\mathbf{X},\mathbf{H})c_{n,i}(\mathbf{X},\mathbf{H})}{S_{i}(\mathbf{X})}\geq D, ~i\in\mathcal{I}(\mathbf{X}),\label{const:ofdma-size-new1}
\end{align}
\begin{align}
&B\log_2\left(1+\frac{p_{n,i}(\mathbf{X},\mathbf{H})H_{n,k}}{n_0}\right)\geq c_{n,i}(\mathbf{X},\mathbf{H}),\nonumber\\
&\quad\quad\quad\quad\quad\quad\quad\quad\quad\quad~ k\in\mathcal{K}_i(\mathbf{X}),i\in\mathcal{I}(\mathbf{X}),\label{const:ofdma-size-new2}
\end{align}
 where} $n_0$ is the power of the complex additive  white Gaussian noise on each subcarrier at each receiver.
\textcolor{black}{Denote} $\bm{\mu}(\mathbf{X},\mathbf{H})\triangleq(\mu_{n,i}(\mathbf{X},\mathbf{H}))_{n\in\mathcal{N},i\in\mathcal{I}(\mathbf{X})}$,
 $\mathbf{p}(\mathbf{X},\mathbf{H})\triangleq(p_{n,i}(\mathbf{X},\mathbf{H}))_{n\in\mathcal{N},i\in\mathcal{I}(\mathbf{X})}$
\textcolor{black}{and $\mathbf{c}(\mathbf{X},\mathbf{H})\triangleq(c_{n,i}(\mathbf{X},\mathbf{H}))_{n\in\mathcal{N},i\in\mathcal{I}(\mathbf{X})}$.}

\vspace{-0.10cm}
\section{Transmission Power Minimization}\label{sec:power}
\vspace{-0.05cm}
\subsection{Problem Formulation}
\vspace{-0.05cm}
\textcolor{black}{Given the video quality (i.e., encoding rate of each tile $D$),
we would like to minimize the transmission power.}
\begin{Prob} [Transmission Power Minimization]\label{ofdma-P1}
For all $(\mathbf{X},\mathbf{H})\in\bm{\mathcal{X}}\times\bm{\mathcal{H}}$,
\begin{align}
P^{\star}(\mathbf{X},\mathbf{H})\triangleq&\min_{\bm{\mu}(\mathbf{X},\mathbf{H}),\mathbf{p}(\mathbf{X},\mathbf{H}),\mathbf{c}(\mathbf{X},\mathbf{H})}~P(\bm{\mu}(\mathbf{X},\mathbf{H}),\bm{p}(\mathbf{X},\mathbf{H}))\nonumber\\
     %s.t.\quad & \ \zeta_{u,l}\geq\Gamma_{l},l\in\Upsilon_{g},u\in\mathcal{U}_{g},g\in\mathcal{G}.\nonumber\\
    \mathrm{s.t.} ~~&\eqref{const:ofdma-integer1},~\eqref{const:ofdma-integer2},~\eqref{const:ofdma-power1},~\eqref{const:ofdma-power0},~\eqref{const:ofdma-size-new1},~\eqref{const:ofdma-size-new2}.\nonumber
\end{align}
Let \textcolor{black}{($\bm{\mu}^{\star}_{\text{e}}(\mathbf{X},\mathbf{H})$, $\mathbf{p}^{\star}_{\text{e}}(\mathbf{X},\mathbf{H})$, $\mathbf{c}^{\star}_{\text{e}}(\mathbf{X},\mathbf{H}))$} denote an optimal solution.
\end{Prob}

Problem~\ref{ofdma-P1} is a mixed discrete-continuous optimization problem.
\textcolor{black}{The number of variables of Problem~\ref{ofdma-P1} (proportional to the number of sets of tiles $I(\mathbf{X})$) is much smaller than that in \cite{multicast-nojoint} (proportional to the number of tiles to be transmitted to all users $|\Phi(\mathbf{X})|$), \textcolor{black}{as $I(\mathbf{X})$ is
much smaller than $|\Phi(\mathbf{X})|$.}}
\vspace{-0.25cm}
\subsection{Optimal Solution}\label{sec:power-solution}
\vspace{-0.05cm}
In this part, we obtain a globally optimal solution of  Problem~\ref{ofdma-P1} (under a mild condition). First,
\textcolor{black}{to reduce computational complexity, we eliminate $\mathbf{c}(\mathbf{X},\mathbf{H})$ and simplify the constraints in \eqref{const:ofdma-power0},
\eqref{const:ofdma-size-new1} and \eqref{const:ofdma-size-new2} to
\begin{align}
&\sum\nolimits_{n\in\mathcal{N}}\frac{\mu_{n,i}(\mathbf{X},\mathbf{H})B}{S_{i}(\mathbf{X})}\log_2\left(1+\frac{p_{n,i}(\mathbf{X},\mathbf{H})H^{\min}_{n,i}(\mathbf{X},\mathbf{H})}{n_0}\right)\nonumber\\
&\geq D,~i\in\mathcal{I}(\mathbf{X}),\label{const:ofdma-size-original}
\end{align}}
where
$H^{\min}_{n,i}(\mathbf{X},\mathbf{H})\triangleq\min_{k\in\mathcal{K}_{i}(\mathbf{X})}H_{n,k}$.
Next, we  \textcolor{black}{relax} the constraints in \eqref{const:ofdma-integer1} to
\begin{align}
&\mu_{n,i}(\mathbf{X},\mathbf{H})\geq 0,~n\in\mathcal{N},~i\in\mathcal{I}(\mathbf{X}). \label{const:ofdma-relax}
\end{align}
Then, \textcolor{black}{let $P_{n,i}(\mathbf{X},\mathbf{H})\triangleq\mu_{n,i}(\mathbf{X},\mathbf{H})p_{n,i}(\mathbf{X},\mathbf{H})$ and $\mathbf{P}(\mathbf{X},\mathbf{H})\triangleq(P_{n,i}(\mathbf{X},\mathbf{H}))_{n\in\mathcal{N},i\in\mathcal{I}(\mathbf{X})}$.}
Thus, Problem~\ref{ofdma-P1} \textcolor{black}{can be} transformed to the following  problem.
\begin{Prob} [Relaxation of Problem~\ref{ofdma-P1}]\label{ofdma-ASP1}
\begin{align}
&\min_{\bm{\mu}(\mathbf{X},\mathbf{H}),\mathbf{P}(\mathbf{X},\mathbf{H})}~\sum\nolimits_{n\in\mathcal{N}}\sum\nolimits_{i\in\mathcal{I}(\mathbf{X})}P_{n,i}(\mathbf{X},\mathbf{H})\nonumber\\
     %s.t.\quad & \ \zeta_{u,l}\geq\Gamma_{l},l\in\Upsilon_{g},u\in\mathcal{U}_{g},g\in\mathcal{G}.\nonumber\\
    &\mathrm{s.t.}~
    \eqref{const:ofdma-integer2},~\eqref{const:ofdma-relax},\nonumber\\
    &P_{n,i}(\mathbf{X},\mathbf{H})\geq 0,~i\in\mathcal{I}(\mathbf{X}),n\in\mathcal{N},\label{const:ofdma-power2}\\
        &\sum\nolimits_{n\in\mathcal{N}}\frac{\mu_{n,i}(\mathbf{X},\mathbf{H})B}{   S_{i}(\mathbf{X})}\log_2\left(1+\frac{P_{n,i}(\mathbf{X},\mathbf{H})H^{\min}_{n,i}(\mathbf{X},\mathbf{H})}{\mu_{n,i}(\mathbf{X},\mathbf{H})n_0}\right)\nonumber\\
    &\geq D,~i\in\mathcal{I}(\mathbf{X}).\label{const:ofdma-data size2}
\end{align}
%Let ($\bm{\mu}^{\dagger}_{\text{e}}$, $\mathbf{P}^{\dagger}_{\text{e}}$) denote an optimal solution.
\end{Prob}

\textcolor{black}{Problem~\ref{ofdma-ASP1} is convex and} can be solved using \textcolor{black}{KKT conditions~\cite{1999OFDMA}.}
Let $\lambda_{i}(\mathbf{X},\mathbf{H}),i\in\mathcal{I}(\mathbf{X})$ denote the lagrange multipliers \textcolor{black}{with respect to} the constraints in \eqref{const:ofdma-data size2}.
Define
\begin{align}
&f_{n,i}(\mathbf{X},\mathbf{H},\lambda_i)\triangleq \left[\frac{B\lambda_{i}}{S_{i}(\mathbf{X})\ln2}-\frac{n_0}{H^{\min}_{n,i}(\mathbf{X},\mathbf{H})}\right]^{+},\nonumber
\end{align}
\begin{align}
&W_{n,i}(\mathbf{X},\mathbf{H},\lambda_i)\triangleq\nonumber\\
&\frac{\lambda_{i}B}{S_{i}(\mathbf{X})}
\left(\log_2\left(1
+\frac{H^{\min}_{n,i}(\mathbf{X},\mathbf{H})f_{n,i}(\mathbf{X},\mathbf{H},\lambda_i)}{n_0}\right)
\right.\nonumber\\
&~~~~~~~~~~\left.-\frac{H^{\min}_{n,i}(\mathbf{X},\mathbf{H})f_{n,i}(\mathbf{X},\mathbf{H},\lambda_i)}{\left(n_0+H^{\min}_{n,i}(\mathbf{X},\mathbf{H})f_{n,i}(\mathbf{X},\mathbf{H},\lambda_i)\right)\ln2}\right).\nonumber
\end{align}
\begin{Lem}[Optimal Solution of Problem~\ref{ofdma-P1}]\label{lem:opt solution}
Suppose that for all $n\in\mathcal{N}$, there exists a unique $i_n$ such that \textcolor{black}{$W_{n,i_n}(\mathbf{X},\mathbf{H},\lambda_{i_n})=\max_{j\in\mathcal{I}(\mathbf{X})}W_{n,j}(\mathbf{X},\mathbf{H},\lambda_j)$}. Then, ($\bm{\mu}^{\star}_{\text{e}}(\mathbf{X},\mathbf{H})$, $\mathbf{p}^{\star}_{\text{e}}(\mathbf{X},\mathbf{H})$, $\mathbf{c}^{\star}_{\text{e}}(\mathbf{X},\mathbf{H})$)  is identical to the optimal solution of Problem~\ref{ofdma-ASP1}, \textcolor{black}{where} for all $n\in\mathcal{N}$ and $i\in\mathcal{I}(\mathbf{X})$,
\begin{align}
&\mu_{{\text{e}},n,i}^{\star}(\mathbf{X},\mathbf{H})=
             \begin{cases}
             1, ~ i=\argmax\limits_{j\in\mathcal{I}(\mathbf{X})}W_{n,j}(\mathbf{X},\mathbf{H},\lambda_j^{\star}(\mathbf{X},\mathbf{H})), \\
             0, ~ \text{otherwise},
             \end{cases}\nonumber\\
&p_{{\text{e}},n,i}^{\star}(\mathbf{X},\mathbf{H})=\mu^{\star}_{{\text{e}},n,i}(\mathbf{X},\mathbf{H})
f_{n,i}(\mathbf{X},\mathbf{H},\lambda_i^{\star}(\mathbf{X},\mathbf{H})),\nonumber\\
&\textcolor{black}{c^{\star}_{{\text{e}},n,i}(\mathbf{X},\mathbf{H})=B\log_2\left(1+\frac{p_{{\text{e}},n,i}^{\star}(\mathbf{X},\mathbf{H})H^{\min}_{n,i}(\mathbf{X},\mathbf{H})}{n_0}\right)}.\nonumber
\end{align}
\textcolor{black}{Here,} $\lambda_{i}^{\star}(\mathbf{X},\mathbf{H})$ satisfies
\small{$$\sum\limits_{n\in\mathcal{N}}\frac{\mu^{\star}_{\text{e},n,i}(\mathbf{X},\mathbf{H})B}{S_{i}(\mathbf{X})}\log_2\left(1+\frac{p^{\star}_{\text{e},n,i}(\mathbf{X},\mathbf{H})H^{\min}_{n,i}(\mathbf{X},\mathbf{H})}{n_0}\right)=D.$$}
\end{Lem}

\textcolor{black}{$\lambda_{i}(\mathbf{X},\mathbf{H}),~i\in\mathcal{I}(\mathbf{X})$}
can be obtained using  the subgradient method.
\textcolor{black}{Note that
all previous works on power minimization for multicast in OFDMA systems provide only
low-complexity suboptimal solutions.}

By carefully exploring structural properties of Problem~\ref{ofdma-P1} and Problem~\ref{ofdma-ASP1}, we have the following result.
\begin{Lem}[Optimal Value of Problem~\ref{ofdma-P1}]\label{lem:bound}
\textcolor{black}{(i)} \small{$P^{\star}(\mathbf{X},\mathbf{H})\in\left[\frac{n_0TN}{\max \mathcal{H}}\left(2^{\frac{D\sum\limits_{i\in\mathcal{I}(\mathbf{X})}S_i(\mathbf{X})}{BN}}-1\right),\frac{n_0TI(\mathbf{X})}{\min\mathcal{H}}\left(2^{\frac{D\sum\limits_{i\in\mathcal{I}(\mathbf{X})}S_i(\mathbf{X})}{B}}-1\right)\right]$}.
\normalsize{\textcolor{black}{(ii)} $gP^{\star}(\mathbf{X},\mathbf{H})=P^{\star}(\mathbf{X},\frac{1}{g}\mathbf{H})$, for all $g>0$.}
\end{Lem}

\textcolor{black}{Lemma~\ref{lem:bound} indicates that  the minimum transmission power $P^{\star}(\mathbf{X},\mathbf{H})$ increases  \textcolor{black}{exponentially} with  \textcolor{black}{the total number of tiles to be transmitted, i.e., $\sum\nolimits_{i\in\mathcal{I}(\mathbf{X})}S_i(\mathbf{X})$, approximately, and is inversely proportional to the channel \textcolor{black}{powers}, i.e., $H_k,k\in\mathcal{K}$.}} Note that $\sum\nolimits_{i\in\mathcal{I}(\mathbf{X})}S_i(\mathbf{X})$ reflects the concentration of the viewing  directions of all users.
A smaller value of $\sum\nolimits_{i\in\mathcal{I}(\mathbf{X})}S_i(\mathbf{X})$ means closer viewing directions of all users.

\vspace{-0.32cm}
\section{Quality Maximization}\label{sec:quality}
\vspace{-0.14cm}
\subsection{Problem Formulation}
\vspace{-0.09cm}

Let $\bar{P}$ denote the  transmission  power budget of the system.  Consider the maximum transmission power  constraint:
\begin{align}\label{const:ofdma-power constraint}
P(\bm{\mu}(\mathbf{X},\mathbf{H}),\bm{p}(\mathbf{X},\mathbf{H}))\leq \Bar{P},~(\mathbf{X},\mathbf{H})\in\bm{\mathcal{X}}\times\bm{\mathcal{H}}.
\end{align}
To guarantee user experience, the encoding rate should not change as frequently as the viewing  directions and channel states, and should remain constant within a certain time duration.
Given the transmission power budget $\bar{P}$, we would like to maximize the received video quality (i.e., encoding rate of each tile $D$). Denote $\bm{\mu}\triangleq\left(\bm{\mu}(\mathbf{X},\mathbf{H})\right)_{(\mathbf{X},\mathbf{H})\in\bm{\mathcal{X}}\times\bm{\mathcal{H}}}$,
$\mathbf{p}\triangleq\left(\mathbf{p}(\mathbf{X},\mathbf{H})\right)_{(\mathbf{X},\mathbf{H})\in\bm{\mathcal{X}}\times\bm{\mathcal{H}}}$
and $\mathbf{c}\triangleq\left(\mathbf{c}(\mathbf{X},\mathbf{H})\right)_{(\mathbf{X},\mathbf{H})\in\bm{\mathcal{X}}\times\bm{\mathcal{H}}}$.
\begin{Prob} [Received Video Quality Maximization]\label{ofdma-P2}
\begin{align}
 &D^{\star}_{\text{q}}\triangleq\max_{D,\bm{\mu},\mathbf{p},\mathbf{c}}D\nonumber\\
     %s.t.\quad & \ \zeta_{u,l}\geq\Gamma_{l},l\in\Upsilon_{g},u\in\mathcal{U}_{g},g\in\mathcal{G}.\nonumber\\
    &\mathrm{s.t.} ~~\eqref{const:ofdma-integer1},~\eqref{const:ofdma-integer2},~\eqref{const:ofdma-power1},~\eqref{const:ofdma-power0},~\eqref{const:ofdma-size-new1},~\eqref{const:ofdma-size-new2},~\eqref{const:ofdma-power constraint}.\nonumber
\end{align}
Let ($D_{\text{q}}^{\star}$, $\bm{\mu}_{\text{q}}^{\star}$, $\mathbf{p}_{\text{q}}^{\star}$, $\mathbf{c}_{\text{q}}^{\star}$) denote an optimal solution.
\end{Prob}

Similar to Problem~\ref{ofdma-P1}, Problem~\ref{ofdma-P2} is a mixed discrete-continuous optimization problem.

\vspace{-0.35cm}
\subsection{\textcolor{black}{Near Optimal \textcolor{black}{Solution}}}
\vspace{-0.1cm}
In this part, we obtain a near optimal solution of Problem~\ref{ofdma-P2}.
Let $\mathbf{H}_{\text{min}}$ \textcolor{black}{denote the vector with all $K$ elements being} $\min\mathcal{H}$.
First, we consider a related problem.
\begin{Prob}[Equivalent Problem of Problem~\ref{ofdma-P2}]\label{OFDMA-EP2}
\begin{align}
&\min\nolimits_{\mathbf{X}\in\bm{\mathcal{X}}}D^{\star}_{\text{q}}(\mathbf{X},\mathbf{H}_{\min})\nonumber
\end{align}
where $D^{\star}_{\text{q}}(\mathbf{X},\mathbf{H}_{\min})$ is given by the following subproblem.
\end{Prob}
\begin{Prob} [Subproblem of Problem~\ref{OFDMA-EP2}]\label{OFDMA-SEP2}
For all $\mathbf X\in\bm{\mathcal X}$,
\begin{align}
&D^{\star}_{\text{q}}(\mathbf{X},\mathbf{H}_{\min})\triangleq\max_{D,\{N_i(\mathbf{X},\mathbf{H}_{\min})\}_{i\in\mathcal{I}(\mathbf{X})}}D\nonumber\\
     %s.t.\quad & \ \zeta_{u,l}\geq\Gamma_{l},l\in\Upsilon_{g},u\in\mathcal{U}_{g},g\in\mathcal{G}.\nonumber\\
    &\mathrm{s.t.} \quad N_i(\mathbf{X},\mathbf{H}_{\min})\in\mathcal{N},i\in\mathcal{I}(\mathbf{X}),\label{const:Ni}\\
    &~~~\quad\sum\nolimits_{i\in\mathcal{I}(\mathbf{X})}N_i(\mathbf{X},\mathbf{H}_{\min})\leq N,\label{const:Ni2}\\
    &\sum\limits_{i\in\mathcal{I}(\mathbf{X})}\frac{N_i(\mathbf{X},\mathbf{H}_{\min})n_0}{\min\mathcal{H}}\left(2^{\frac{DS_i(\mathbf{X})}{BN_i(\mathbf{X},\mathbf{H}_{\min})}}-1\right)\leq \Bar{P}.\label{const:ofdma-power constraint2}
\end{align}
\textcolor{black}{Let $(D^{\star}_{\text{q}}(\mathbf{X},\mathbf{H}_{\min}),(N_i^{\star}(\mathbf{X},\mathbf{H}_{\min}))_{i\in\mathcal{I}(\mathbf{X})})$ denote an optimal solution.}
\end{Prob}

\textcolor{black}{Note that $N_i(\mathbf{X},\mathbf{H}_{\min})$ indicates the number of subcarriers assigned to  transmit the symbols for the tiles in $\mathcal{S}_i(\mathbf{X})$}
\textcolor{black}{
at system channel state $\mathbf{H}_{\min}$.}
By carefully exploring structural properties of Problem~\ref{ofdma-P2}, we have the following result.
\begin{Lem}[Equivalence between Problem~\ref{ofdma-P2} and Problem~\ref{OFDMA-EP2}]\label{lem:equivalent}
\textcolor{black}{(i)} \textcolor{black}{The optimal values of  Problem~\ref{ofdma-P2} and Problem~\ref{OFDMA-EP2} are equivalent, i.e., $D^{\star}_{\text{q}}=\min\nolimits_{\mathbf{X}\in\bm{\mathcal{X}}}D^{\star}_{\text{q}}(\mathbf{X},\mathbf{H}_{\min})$.}
\textcolor{black}{(ii)} For all $i\in\mathcal{I}(\mathbf{X}),\mathbf{X}\in\bm{\mathcal{X}}$, $N_i^{\star}(\mathbf{X},\mathbf{H}_{\min})=\sum_{n=1}^{N}\mu_{\text{q},n,i}^{\star}(\mathbf{X},\mathbf{H}_{\min})$,
\begin{align}
&p_{\text{q},n,i}^{\star}(\mathbf{X},\mathbf{H}_{\min})=\nonumber\\
&\begin{cases}
             \frac{n_0}{\min\mathcal{H}}\left(2^{\frac{S_{i}(\mathbf{X})D_{\text{q}}^{\star}(\mathbf{X},\mathbf{H}_{\min})}{BN_{i}^{\star}(\mathbf{X},\mathbf{H}_{\min})}}-1\right),~\mu_{\text{q},n,i}^{\star}(\mathbf{X},\mathbf{H}_{\min})=1;\\
             0, ~\quad\quad\quad\quad\quad\quad\quad\quad\quad\quad\quad\quad~\text{otherwise}.
             \end{cases}\label{lem3}
\end{align}
\end{Lem}

\textcolor{black}{\begin{proof}(sketch)
We eliminate $\mathbf{c}$, and replace \eqref{const:ofdma-power0}, \eqref{const:ofdma-size-new1} and \eqref{const:ofdma-size-new2} with \eqref{const:ofdma-size-original}.
By (ii) of Lemma~\ref{lem:bound},
it is equivalent to consider only $\mathbf{H}_{\min}$ instead of all $\mathbf{H}\in\bm{\mathcal{H}}$.
By
Lemma~\ref{lem:opt solution}, we have $p_{\text{e},n,i}^{\star}(\mathbf{X},\mathbf{H}_{\min})= p_{\text{e},m,i}^{\star}(\mathbf{X},\mathbf{H}_{\min})$,
for all $n,m$ with $\mu_{\text{e},n,i}^{\star}(\mathbf{X},\mathbf{H}_{\min})=\mu_{\text{e},m,i}^{\star}(\mathbf{X},\mathbf{H}_{\min})=1$.
By
 \eqref{const:ofdma-size-original} and \eqref{const:ofdma-power constraint},
we can show \eqref{lem3} by contradiction.
Substituting \eqref{lem3} into \eqref{const:ofdma-power constraint},
we can obtain \eqref{const:ofdma-power constraint2}.
By setting $N_i(\mathbf{X},\mathbf{H}_{\min})=\sum_{n=1}^{N}\mu_{n,i}(\mathbf{X},\mathbf{H}_{\min})$,
\eqref{const:ofdma-integer1} and \eqref{const:ofdma-integer2} can be transformed to \eqref{const:Ni} and \eqref{const:Ni2}.
Thus,  Problem~\ref{ofdma-P2} can be equivalently transformed to Problem~\ref{OFDMA-EP2}.
\end{proof}}

%\vspace{-1.2cm}
Relaxing \eqref{const:Ni} to
$N_i(\mathbf{X},\mathbf{H}_{\min})\in\left[1,N\right],~i\in\mathcal{I}(\mathbf{X})$,
Problem~\ref{OFDMA-SEP2} can be transformed to a convex problem, which can be solved using KKT conditions.
\begin{Lem}[Optimal Solution of  Relaxation of Problem~\ref{OFDMA-SEP2}]\label{lem:opt-relaxP5}
The optimal solution of  the relaxation of Problem~\ref{OFDMA-SEP2} is
\begin{align}
&N^{\dagger}_i(\mathbf{X},\mathbf{H}_{\min})=\frac{S_i(\mathbf{X})N}{\sum_{i\in\mathcal{I}(\mathbf{X})}S_i(\mathbf{X})},~i\in\mathcal{I}(\mathbf{X}),\nonumber
\end{align}
\begin{align}
&D^{\dagger}(\mathbf{X},\mathbf{H}_{\min})=\frac{BN\ln(\frac{\Bar{P}\min\mathcal{H}}{Nn_0}+1)}{\ln{2}\sum\nolimits_{i\in\mathcal{I}(\mathbf{X})}S_{i}(\mathbf{X})}.\nonumber
\end{align}
\end{Lem}

\textcolor{black}{Lemma~\ref{lem:opt-relaxP5} indicates that
$N^{\dagger}_i(\mathbf{X},\mathbf{H}_{\min})$ is proportional to $S_i(\mathbf{X})$.}
By  exploring  properties of  Problem~\ref{ofdma-P2}, Problem~\ref{OFDMA-EP2} and Problem~\ref{OFDMA-SEP2} \textcolor{black}{and by Lemma~\ref{lem:opt-relaxP5}}, we have the following result.
\begin{Lem}[Optimal Value of Problem~\ref{ofdma-P2}]\label{lem:quality opt structure}
$D_{\text{q}}^{\star}\in\left[\frac{B\ln(\frac{\Bar{P}\min\mathcal{H}}{\max_{\mathbf{X}\in\bm{\mathcal{X}}}I(\mathbf{X})n_0}+1)}{\ln{2}\max\nolimits_{\mathbf{X}\in\bm{\mathcal{X}}}\sum\nolimits_{i\in\mathcal{I}(\mathbf{X})}S_{i}(\mathbf{X})},\frac{BN\ln(\frac{\Bar{P}\min\mathcal{H}}{Nn_0}+1)}{\ln{2}\max\nolimits_{\mathbf{X}\in\bm{\mathcal{X}}}\sum\nolimits_{i\in\mathcal{I}(\mathbf{X})}S_{i}(\mathbf{X})}\right]$.
\end{Lem}

Lemma~\ref{lem:quality opt structure} indicates that approximately,   $D_{\text{q}}^{\star}$ \textcolor{black}{is affected by the smallest channel power $\min\mathcal{H}$ among all channel powers}, and is inversely proportional to $\max\nolimits_{\mathbf{X}\in\bm{\mathcal{X}}}\sum\nolimits_{i\in\mathcal{I}(\mathbf{X})}S_i(\mathbf{X})$ which represents the maximum number of tiles that need to be transmitted for all viewing directions.

\textcolor{black}{Now, we propose a low complexity algorithm, i.e., Algorithm~\ref{Greedy Algorithm}, to  obtain a near optimal encoding rate of each tile  of Problem~\ref{ofdma-P2}, denoted by $D_{\text{q}}^{\diamond}$,
 \textcolor{black}{by constructing a feasible solution based on the solution in Lemma~\ref{lem:opt-relaxP5}
 in a greedy manner.}}
\begin{algorithm}[h]
    \caption{Near Optimal Solution of Problem~\ref{ofdma-P2}}
%Input: $C_{u}$, $\Gamma_{l}$, $u\in\mathcal{U}_{g},l\in\Upsilon_{g}$, $g\in\{1,2\}$;\\
%Output: the optimal rank-one solution  $(X_{1},X_{2})$;\\
    %\begin{multicols}{2}
    \begin{footnotesize}
     \begin{algorithmic}[1]
 %          \STATE Solve Problem~\ref{OFDMA-SEP2} with continuous $N_i(\mathbf{X},\mathbf{H}_{\min})\in(1,N),i\in\mathcal{I}(\mathbf{X})$ using KKT conditions;
           \STATE For all $i\in\mathcal{I}(\mathbf{X})$, set $N_i^{\diamond}(\mathbf{X},\mathbf{H}_{\min})=\lfloor N_i^{\dagger}(\mathbf{X},\mathbf{H}_{\min})\rfloor$;
           \WHILE{$\sum_{i\in\mathcal{I}(\mathbf{X})}N_{i}^{\diamond}(\mathbf{X},\mathbf{H}_{\min})<N$}
%           \STATE Compute $j=\argmax_{i\in\mathcal{I}(\mathbf{X})} \frac{S_{i}(\mathbf{X})\ln2}{B} 2^{\frac{S_i(\mathbf{X})}{BN_i(\mathbf{X},\mathbf{H}_{\min})}}$;
           \STATE Set $N^{\diamond}_{i^{\star}}(\mathbf{X},\mathbf{H}_{\min})=N^{\diamond}_{i^{\star}}(\mathbf{X},\mathbf{H}_{\min})+1$, where $i^{\star}=\argmax_{i\in\mathcal{I}(\mathbf{X})} \frac{S_{i}(\mathbf{X})\ln2}{B} 2^{\frac{S_i(\mathbf{X})}{BN^{\diamond}_i(\mathbf{X},\mathbf{H}_{\min})}}$;
           \ENDWHILE
           \STATE For all $\mathbf{X}\in\bm{\mathcal{X}}$, obtain $D^{\diamond}_{\text{q}}(\mathbf{X},\mathbf{H}_{\min})$ by solving  $\sum\limits_{i\in\mathcal{I}(\mathbf{X})}\frac{N^{\diamond}_i(\mathbf{X},\mathbf{H}_{\min})n_0}{\min\mathcal{H}}\left(2^{\frac{D^{\diamond}_{\text{q}}(\mathbf{X},\mathbf{H}_{\min})S_i(\mathbf{X})}{BN^{\diamond}_i(\mathbf{X},\mathbf{H}_{\min})}}-1\right)=\Bar{P}$ using bisection search;
           \STATE Set $D^{\diamond}_{\text{q}}=\min\nolimits_{\mathbf{X}\in\bm{\mathcal{X}}}D^{\diamond}_{\text{q}}(\mathbf{X},\mathbf{H}_{\min})$.
    \end{algorithmic}\label{Greedy Algorithm}
    \end{footnotesize}
    %\end{multicols}
    \vspace{-0.1cm}
\end{algorithm}
\vspace{-0.65cm}
\section{Simulation}\label{sec:simulation}
\vspace{-0.18cm}
In this section, we compare the proposed solutions in Section~\ref{sec:power} and Section~\ref{sec:quality} with two baselines using numerical results.
\textcolor{black}{Baseline 1 considers serving each user separately using unicast in an optimal way, similar to the \textcolor{black}{proposed solutions}. Baseline~2 \textcolor{black}{considers  multicast  but with} equal subcarrier allocation for each transmitted tile and optimal \textcolor{black}{transmission power as well as transmission rate allocation}  based on the equal subcarrier allocation.}
In the simulation, we use Kvazaar as the 360 VR video encoder and video sequence $Boxing$ as the video source.
To avoid view switch delay, we transmit
extra $15^{\circ}$ in the four directions of each requested FoV \textcolor{black}{\cite{7532381}}.
\textcolor{black}{Different 360 VR videos in general have different popularity distributions for
viewing directions. To illustrate the importance of the concentration of the viewing directions,} we assume all users randomly and independently select viewing directions according to Zipf distribution\footnote{\textcolor{black}{Zipf distribution is widely used to model content popularity in Internet and wireless
networks. For any popularity rank,
a larger Zipf exponent $\gamma$ indicates
a smaller tail of the popularity distribution, i,e, higher concentration of
requests for contents. Here, we adopt Zipf distribution for ease of exposition.}} for the $M_h\times M_v$ viewing directions.
\textcolor{black}{In particular, suppose viewing direction $(m_h,m_v)$ is of rank $(m_h-1)M_v+m_v$ and $\Pr[\mathbf{X}_k=(m_h,m_v)]=\frac{((m_h-1)M_v+m_v)^{-\gamma}}{\sum\nolimits_{i=1,\ldots,M_hM_v}i^{-\gamma}}$, where $\gamma$ is the Zipf exponent.
In addition, assume
$\sqrt{H_{n,k}}$, \textcolor{black}{$n\in\mathcal{N}$, $k\in\mathcal{K}$} are randomly and independently distributed according to $\mathcal{CN}(0,\frac{1}{d})$, where $d$ reflects the path loss. We randomly choose 100 global channel states to form $\bm{\mathcal{H}}$, and evaluate the average transmission power over $\bm{\mathcal{H}}$.}

\begin{figure}[t]
\vspace*{-1.2cm}
\begin{center}
\vspace*{-0.07cm}
 \subfigure[\small{Average transmission power versus $\gamma$. $K=3$, $D=30$ kbit/s, \textcolor{black}{$d=10^{3}$}.}]
 {\resizebox{3.3cm}{!}{\includegraphics{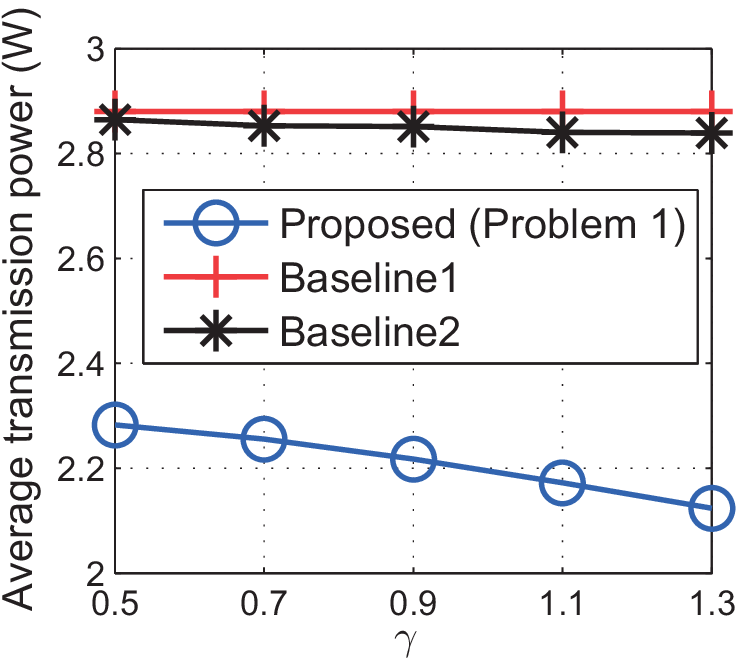}}}
 \vspace*{-0.07cm}
  \subfigure[\small{Encoding rate of each tile versus $\gamma$. $K=4$, $d=600$, \textcolor{black}{$\Bar{P}=10^4$ W}.}]
 {\resizebox{3.3cm}{!}{\includegraphics{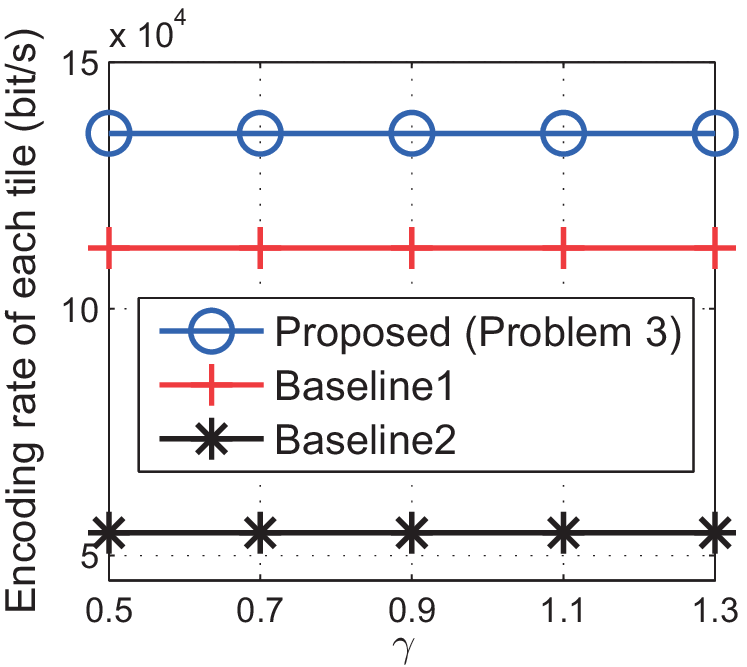}}}
 \end{center}
 \vspace*{-0.37cm}
   \caption{\small{Performance comparison. $F_h=F_v=100^{\circ}$, \textcolor{black}{$M_h\times M_v=30\times2$}, $V_h\times V_v=30\times15$, $B =39$ kHz, $N=128$, $n_0 = 10^{-9}$ W.}}
   \label{fig:power}
\vspace*{-0.60cm}
\end{figure}

Fig.~\ref{fig:power} (a) illustrates the average transmission power versus the Zipf exponent $\gamma$.
\textcolor{black}{We can see that  the average transmission
power of each multicast scheme  decreases with $\gamma$, as multicast
opportunities increase with $\gamma$;
the average transmission power of the unicast scheme almost does not change  with $\gamma$, as it does not capture multicast opportunities.
Fig.~\ref{fig:power} (b) illustrates the encoding rate of each tile versus
 $\gamma$.
We can see that  the encoding rate of each tile of each
 scheme does not change with $\gamma$,
as that of each multicast scheme is determined by $\argmax_{\mathbf{X}\in\bm{\mathcal{X}}}\sum_{i\in\mathcal{I}(\mathbf{X})}S_i(\mathbf{X})$
corresponding to the case with the fewest multicast opportunities, and that of the
unicast scheme does not depend on $\mathbf{X}$.}
From Fig.~\ref{fig:power}, we can also observe that the
proposed solutions outperform the two baselines. Specifically,
the gains of the proposed solutions over Baseline~1 arise from
the fact that the proposed solutions utilize multicast. The gains
of the proposed solutions over Baseline~2 are due to the fact
that the proposed solutions carefully allocate  \textcolor{black}{subcarrier, transmission power and
transmission rate.}

\vspace{-0.35cm}
\section{Conclusion}
\vspace{-0.14cm}
In this letter, we \textcolor{black}{studied} optimal multicast of tiled 360 VR video
in an  OFDMA system, and \textcolor{black}{formulated} two non-convex
optimization problems, i.e., the minimization of the average
transmission power for given video
quality, and the maximization of the received video quality
 for given
\textcolor{black}{transmission power budget}. We  \textcolor{black}{obtained}  a globally optimal closed-form solution and a near optimal solution of the two non-convex problems, separately, and \textcolor{black}{revealed} important design insights for tiled 360 VR multicast.
\textcolor{black}{This letter opens up several directions for future research. For instance, the proposed multicast
mechanism and optimization framework can be extended
to design optimal multi-quality multicast of tiled 360 VR video in OFDMA systems.
In addition, a possible direction for future research is to design optimal
single-quality or multi-quality  multicast of tiled 360 VR video in different wireless systems.}

\vspace{-0.37cm}
\bibliographystyle{IEEEtran}

\end{document}